\documentclass[prd,groupedaddress, showpacs, doublecolumn,nofootinbib]{revtex4} %
\usepackage{graphicx}
\usepackage{dcolumn}
\usepackage{bm}
\usepackage[hyperfootnotes=true]{hyperref}
\usepackage{amssymb}
\usepackage{amsmath}
\usepackage{epsfig}
\usepackage{hhline}
\usepackage{color}
\usepackage{multirow}
\usepackage{cancel}

\def\lapp{\mathrel{\rlap{\raise.5ex\hbox{$<$}}
                    {\lower.5ex\hbox{$\sim$}}}}
\def\gapp{\mathrel{\rlap{\raise.5ex\hbox{$>$}}
                    {\lower.5ex\hbox{$\sim$}}}}

{\newcommand{\lsim}{\mbox{\raisebox{-.6ex}{~$\stackrel{<}{\sim}$~}}}
{
\newcommand{\bmt}{\begin{pmatrix}}
\newcommand{\emt}{\end{pmatrix}}
\newcommand{\ba}{\begin{array}{c}}
\newcommand{\ea}{\end{array}}
\newcommand{\be}{\begin{equation}}
\newcommand{\ee}{\end{equation}}
\newcommand{\bea}{\begin{eqnarray}}
\newcommand{\eea}{\end{eqnarray}}

\newcommand{\bi}{\begin{itemize}}
\newcommand{\ei}{\end{itemize}}

\newcommand{\baz}{\begin{array}{cc}}

\newcommand{\mathsym}[1]{{}}

\newcommand{\bt}{\begin{tabular}}
\newcommand{\et}{\end{tabular}}

\newcommand{\benu}{\begin{enumerate}}
\newcommand{\eenu}{\end{enumerate}}
\newcommand{\bav}{\begin{array}{cccc}}

\begin{document}
\title{\bf Dark matter assisted Dirac leptogenesis and neutrino mass}

\author{Nimmala Narendra}
\email{ph14resch01002@iith.ac.in}
%\affiliation{Department of Physics, Indian Institute of Technology,
 % Hyderabad, Kandi, Sangareddy, 502285, Telangana, India}
\author{Nirakar Sahoo}
\email{nirakar.pintu.sahoo@gmail.com}
\author{Narendra Sahu}
\email{nsahu@iith.ac.in}
\affiliation{Department of Physics, Indian Institute of Technology,
  Hyderabad, Kandi, Sangareddy, 502285, Telangana, India}

\begin{abstract}
We propose an extension of the standard model with $U(1)_{B-L} \times Z_{2}$ symmetry. In this model by assuming that 
the neutrinos are Dirac ({\it i.e.} $B-L$ is an exact symmetry), we found a simultaneous solution for non zero neutrino masses 
and dark matter content of the universe. The observed baryon asymmetry of the universe is also explained using Dirac Leptogenesis, 
which is assisted by a dark sector, gauged under a $U(1)_D$ symmetry. The latter symmetry of the dark sector is broken at a TeV scale 
and thereby giving mass to a neutral gauge boson $Z_D$. The standard model Z-boson mixes with the gauge boson $Z_D$ at one loop 
level and paves a way to detect the dark matter through spin independent elastic scattering at terrestrial laboratories. 
\end{abstract}

\maketitle

\newpage
\section{Introduction} \label{Intro}
The standard model (SM), which is based on the gauge group $SU(3)_C\times SU(2)_L\times U(1)_{Y}$, is a successful 
theory of fundamental particles of nature and their interactions. After the Higgs discovery, it seems to be complete. However, 
there are many unsolved issues which are not addressed within the framework of SM. In particular, the non-zero neutrino 
masses, baryon asymmetry of the Universe, existence of dark matter {\it etc.}. These problems beg for a successful theory in 
physics beyond the SM. 

The observed galactic rotation curve, gravitational lensing and large scale structure of the Universe collectively hint towards 
the existence of an invisible matter, called dark matter. In fact, the relic abundance of dark matter has been precisely 
determined by the satellite based experiments, such as WMAP~\cite{Hinshaw:2012aka} and PLANCK~\cite{Ade:2015xua} to be $\Omega_{\rm DM}h^2
=0.1199 \pm 0.0027$. Hitherto the existence of dark matter is shown in a larger scale ($\gtrsim$ a few kpc) only via its gravitational 
interaction. However, the particle nature of dark matter is remained elusive till today and needs to be explored in a framework of 
physics beyond the SM.   

Within the SM, the neutrinos are exactly massless. This can be traced to a conserved $B-L$ symmetry within the SM, where $B$ and $L$ 
stands for net baryon and lepton number respectively. However, the oscillation experiments~\cite{solar-expt, atmos-expt, kamland} have 
successfully demonstrated that the neutrinos have sub-eV masses. One attractive way to explain the small masses of active neutrinos is to 
introduce the lepton number violation by two units through the dimension five operator $\ell \ell HH/\Lambda$~\cite{Weinberg:1979sa}, where 
$\ell, H$ are the lepton and Higgs doublet respectively and $\Lambda$ is the scale at which the new physics is expected to arise. After 
electroweak phase transition, the neutrinos acquire a Majorana mass of the order $m_\nu = \langle H \rangle^2/\Lambda$. Naively this 
implies that the sub-eV masses of neutrinos indicate the scale of new physics to be $\Lambda \sim {\cal O}(10^{14}){\rm GeV}$. Note 
that the effective dimension-5 operator can be realized in many extensions of the standard model, the so called seesaw 
mechanisms~\cite{type1_seesaw, type2_seesaw, type3_seesaw}. In these models, the mass scale of new particles is expected to be at a 
scale of $\Lambda$. Therefore, it is imagined that in the early Universe, when the temperature of thermal bath is high enough, namely 
$T\gtrsim \Lambda$, the lepton number violation can occur rapidly, while it is suppressed today. As a result, a net lepton 
asymmetry~\cite{fukugita.86, baryo_lepto_group} can be generated through CP violating out-of-equilibrium decay~\cite{sakharov.67} of 
these heavy particles at $T \sim \Lambda$, which is then converted to the observed baryon asymmetry of the Universe through the 
electroweak sphaleron transitions. The lepton number violating interactions ($\Delta L = 2$), which also indicate Majorana 
nature of neutrinos, can be proved at ongoing neutrinoless double beta decay experiments~\cite{double_betadecay_expts}. But till now 
there is no positive result found in those experiments. So there is still a chance of hope that the neutrinos might be Dirac in nature. 
In other words, $B-L$ is an exact global symmetry of the SM Lagrangian. 

Even the neutrinos are Dirac in nature (i.e. $B-L$ is exactly conserved), the baryon asymmetry of the Universe must be explained since 
it is an observed fact. It has been explored largely in the name of Dirac leptogenesis~\cite{Dick, Cerdeno:2006ha, Gu:2006dc, Gu:2007mc, 
Murayama:2002je, Thomas:2005rs, Borah:2016zbd}, which connect Dirac mass of neutrinos with the observed baryon asymmetry of the Universe. 
The key point of this mechanism is that the equilibration time between left and right-handed neutrinos mediated via SM Higgs (i.e. 
$Y\overline{\nu_R}H\nu_L$)is much less than the $(B+L)$ violating sphaleron transitions above electroweak phase transition. Therefore, if we demand that $B-L= 
B-(L_{\rm SM}+L_{\nu_R})=0$\cite{Cerdeno:2006ha}, then we see that a net $B-L_{\rm SM}$ is generated in terms of $L_{\nu_R}$. The electroweak sphalerons will 
not act on $L_{\nu_R}$, as $\nu_R$ is singlet under $SU(2)_L$, while the non-zero $B-L_{\rm SM}$ will be converted to a net $B$ asymmetry 
via $B+L$ violating sphaleron transitions. 

\begin{figure} [h!]
\centering
\includegraphics[width=50mm]{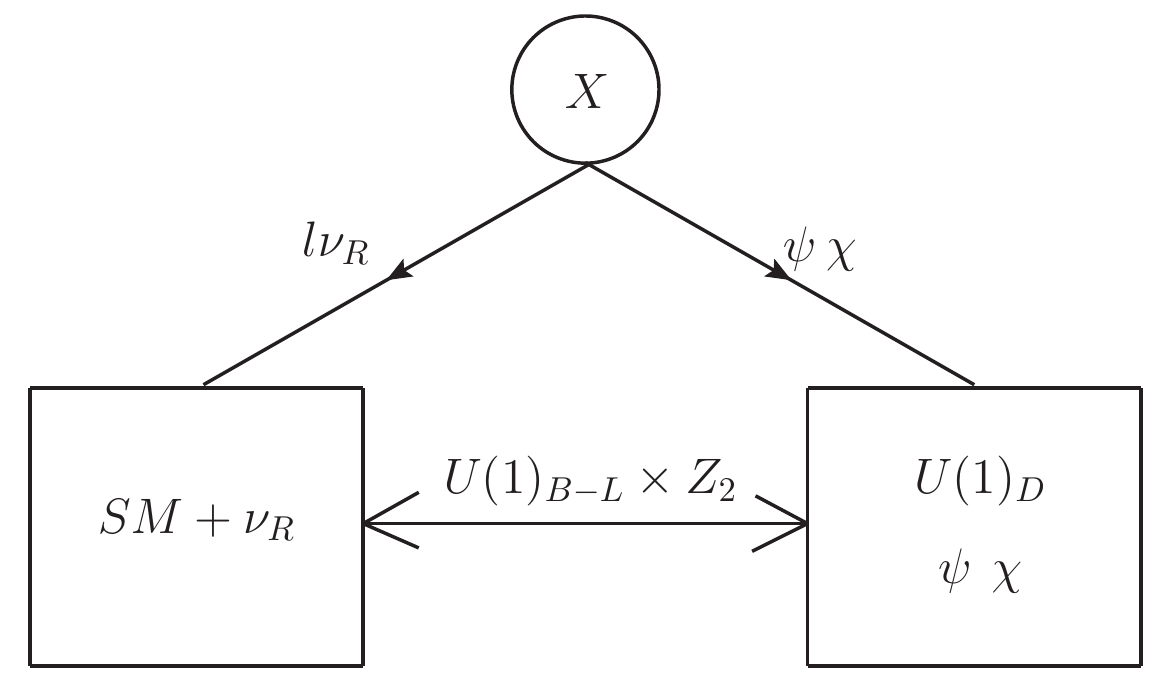}
\caption{\footnotesize{Schematic diagram of a heavy scalar decay to visible and dark sectors, where the ``dark sector" we mean the 
particles being charged under a $U(1)_D$ symmetry and constitutes two vector-like Dirac fermions:  $\psi$ (doublet under $SU(2)_L$) 
and $\chi$ (singlet under $SU(2)_L$). The lightest dark sector particle $\chi$ is a candidate of dark matter.}}
\label{Diagram}
\end{figure}

In this paper we study the consequences of Dirac nature of neutrinos to a simultaneous solution of dark matter and baryon asymmetry 
of the Universe. We extend the SM by introducing a dark sector constituting of two vector-like Dirac fermions: $\psi$, a doublet under 
$SU(2)_L$, and $\chi$, a singlet under $SU(2)_L$, as shown in the Fig.~\ref{Diagram}. See also refs.\cite{Bhattacharya:2015qpa,
Bhattacharya:2017sml}. The dark sector is gauged under a $U(1)_D$ symmetry. An over all symmetry $U(1)_{B-L}\times Z_{2}$ is also 
imposed to ensure that the neutrinos are Dirac and the lightest particle $\chi$ in the dark sector is a candidate of dark matter, being 
odd under the $Z_2$ symmetry. A heavy scalar doublet $X$, odd under the discrete $Z_2$ symmetry, is introduced such that its CP-violating 
out-of-equilibrium decay to $\ell \nu_R$ and $\psi \chi$ can generate equal and opposite lepton asymmetry in both the channels, where $\nu_R$ 
is odd under $Z_2$. The lepton asymmetry produced in the channel $X\to \ell \nu_R$ can be converted to a net baryon asymmetry via $B+L$ 
violating sphaleron transitions, where as the asymmetry in the channel $X\to \psi \chi$ will remain intact as $\psi$ is a vector-like 
Dirac fermion~\cite{Arina:2011cu,Arina:2012fb,Arina:2012aj}. Notice that the $Z_2$ symmetry forbids the term $\overline{\nu_R} H \ell$, though 
allowed by $U(1)_{\rm B-L}$. To generate a Dirac mass of the neutrinos we allow $Z_2$ to break softly by the term $\mu^2 X^\dagger H$. As a 
result we generate Dirac mass of the neutrinos to be $M_\nu \sim  \mu^2 \langle H \rangle /M_X^2$, where $M_X$ is the mass of $X$. The 
soft $Z_2$ breaking also introduces a mixing between the neutral component of the doublet $\psi^0$ and $\chi$. As a result the asymmetry in 
$\psi^0$ and $\psi^\pm$ gets converted to a net $\chi$ asymmetry. However, we will show that the $\chi$ asymmetry is significantly 
smaller than the symmetric component because the latter does not get annihilated efficiently to the SM particles through the mixing between 
the gauge bosons $Z$ and $Z_D$, where $Z_D$ is the gauge boson in the dark sector. As a result the relic of $\chi$ constitutes an admixture of 
dominant symmetric component with a small asymmetric components.    

The paper is organized as follows. In sec.~\ref{The Model}, we discuss the proposed model, while in sec.~\ref{Dirac neutrino mass}, we explain the 
Dirac masses of light neutrinos. A brief description about observed baryon asymmetry and DM abundance is given in sec.~\ref{Obs_asy_and_DM_abun}. 
Section \ref{Production of LA} is devoted to explain baryogenesis via leptogenesis from the decay of heavy particles $X$, while section \ref{Dark matter} 
describes DM abundance from the decay of heavy $X$-particles. We discuss the constraints on model parameters from direct detection of DM in sec. \ref{DD} 
and conclude in sec.~\ref{Conclusion}.

\section{The Model}\label{The Model}
%%%%%%%%%%%%%%%%%%%%%%%%%%%%%%%%%%%%%%%%%%%%%%%%%%%%%%%%%%%%%%%%%%%%%%%%%%%%%
We extend the Standard Model with a dark sector, consisting of two vector-like leptons: $\psi$ and $\chi$, where 
$\psi$ is a doublet and $\chi$ is a singlet under $SU(2)_{L}$. The dark sector is gauged under a $U(1)_D$ symmetry, 
which breaks at TeV scales and give mass to the neutral gauge boson $Z_D$. We also introduce three right handed neutrinos 
$\nu_{{R}_{\alpha}}$, $ \alpha = 1,2,3$ and a heavy scalar doublet $X$. A discrete symmetry $Z_2$ is also introduced under which $X,\nu_{R}$ and 
$\chi$ are odd, while $\psi$ and all other SM particles are even. Under $U(1)_{D}$ symmetry $\psi$ and $\chi$ carry non-trivial quantum numbers. 
As a result the trilinear couplings: $\bar{\psi}X\nu_R$ and $\bar{\ell}X\chi$ are forbidden. Here the singlet fermion, $\chi$ is the 
lightest particle in the dark sector and acts as a candidate of dark matter. An overall $B-L$ global symmetry is also introduced as in 
the case of SM. The $B-L$ symmetry remains unbroken and hence ensures that the neutrinos are Dirac in nature. The CP-violating out-of-equilibrium 
decay of heavy scalar $X$ create asymmetries simultaneously in both lepton and dark matter sectors~\cite{Arina:2011cu}. In the visible sector, the 
decay of $X$ to $\ell$ and $\nu_R$, creates an equal and opposite lepton asymmetry in both left and right-handed sectors. The lepton asymmetry in 
the left-handed sector gets converted to a net baryon asymmetry through $B+L$ violating sphaleron transitions, while the asymmetry in the right-handed 
sector remains unaffected until the temperature falls much below the electroweak phase transition. Note that the coupling $\bar{\ell} \nu_{R} H$ and 
$\bar{\psi} H \chi$ are forbidden due to $Z_{2}$ symmetry. The relevant Lagrangian can be written as: 
\begin{equation}
\mathcal{L}\supset \bar{\psi}i\gamma^{\mu} D_{\mu}\psi + \bar{\chi}i \gamma^{\mu}D'_{\mu}\chi + M_{\psi}\bar{\psi} \psi + M_{\chi} \bar{\chi} \chi 
+\left[ f_{kl} \overline{\ell_{k}} \tilde{X} {\nu_{R}}_{l} + \lambda \overline{\psi} \tilde{X} \chi + {\rm h.c.}\right] - V(H, X)\,,
\end{equation}
where
\begin{eqnarray}
D_{\mu} &=& \partial_{\mu}-i\frac{g}{2}\tau^{i}W^{i}_{\mu}-i\frac{g'}{2}YB_{\mu}-ig_{D}Y_{D}(Z_{D})_{\mu}
 \nonumber\\
D'_{\mu} &=& \partial_{\mu}-ig_{D}Y_{D}(Z_{D})_{\mu}
\end{eqnarray}
%%%%%%%%%%%%%%%%%%%%%%%%%%%%%%%%%%%%%%%%%%%%%%%%%%%%%%%%%%%%%%%%%%%%%%%%%%%%
%(D^{\mu} X_{1})^{\dagger} D_{\mu} X_{1} + (D^{\mu} X_{2})^{\dagger} D_{\mu} X_{2} + \overline{\nu_{R}} i \gamma^{\mu}{D_{\mu}} \nu_{R} +  \overline{\psi} i \gamma^{\mu}{D_{\mu}} \psi + \overline{S} i \gamma^{\mu}{D_{\mu}} S \\ 
%%%%%%%%%%%%%%%%%%%%%%%%%%%%%%%%%%%%%%%%%%%%%%%%%%%%%%%%%%%%%%%%%%%%%%%%%%%%
and
\begin{equation}\label{potential}
V(H, X) = -M^{2}_{H} H^{\dagger} H + M_X^2 X^{\dagger} X + \lambda_{H} (H^{\dagger} H)^{2} + \lambda_X (X^{\dagger} X)^{2} + \lambda_{H X} (H^{\dagger} H) (X^\dagger X)\\
%&+ + \lambda_{H X_{2}} (H^{\dagger} H)(X_{2}^{\dagger} X_{2})\\
%&+\mathop{\sum_{\alpha, \beta = 1,2}} \lambda_{\alpha \beta} \left[(X_{\alpha}^{\dagger}X_{\beta})(X_{\beta}^{\dagger} X_{\alpha}) \right] + 
%\left[ \lambda ({X_{1}^{\dagger}H)})(X_{2}^{\dagger}H) + {\rm h.c.}\right] \,.
\end{equation}
\begin{table}[!h]
\renewcommand{\arraystretch}{}
\caption{Quantum numbers of the new particles under the extended symmetry.}
\label{Table}
\centering
\begin{tabular}{|c|c|c|c|}
\hline
Parameter             & $U(1)_{B-L}$ & $U(1)_{D}$ & $Z_{2}$   \\
\hline
$X=(X^{+},X^{0})^{T}$ &       0      &      0     &    -      \\
%\hline
%$X_{2}=(X_{2}^{+},X_{2}^{0})^{T}$ & 0 & - & 0                \\
\hline
$\nu_{R}$             &      -1      &      0     &    -      \\
\hline
$\psi=(\psi^{0},\psi^{-})^{T}$ & -1  &      1     &    +      \\
\hline
$\chi$                &      -1      &      1     &    -      \\
\hline
\end{tabular}
\end{table}

%%%%%%%%%%%%%%%%%%%%%%%%%%%%%%%%%%%%%%%%%%%%%%%%%%%%%%%%%%%%%%%%%%%%%%%%%%%%
\subsection{Dirac mass of neutrinos}\label{Dirac neutrino mass}
We allow the $Z_2$ symmetry to break softly\cite{McDonald:2007ka, Heeck:2013vha} via:
\begin{equation}
\mathcal{L}_{soft} = -\mu^{2} H^{\dagger}X + {\rm h.c.}
\end{equation}

\begin{figure} [h!]
\centering
\includegraphics[width=50mm]{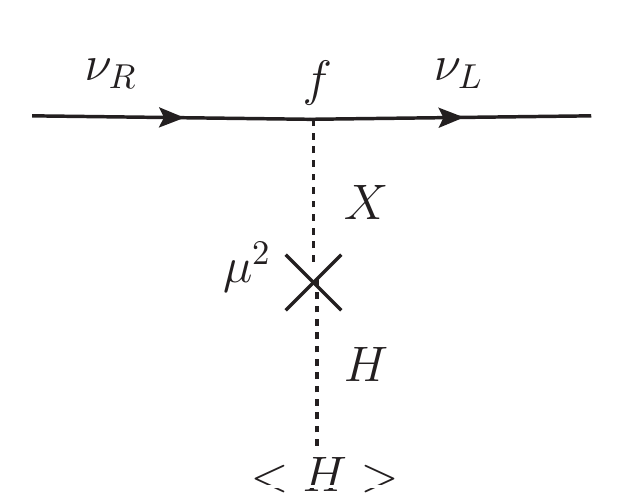}
\caption{\footnotesize{Dirac mass of neutrinos generated via soft $Z_2$ symmetry breaking.}}
\label{Dirac neutrino}
\end{figure}

As a result the Dirac mass of the neutrinos can be generated as shown in the Fig.~\ref{Dirac neutrino}.
After integrating out the heavy field $X$ we get the Dirac neutrino mass:
\begin{equation}
M_\nu = \frac{f \langle H \rangle \mu^{2}}{M_X^{2}}\,,
\end{equation}
where $\langle H \rangle = 174$ GeV, is the Higgs vacuum expectation value. To generate Dirac masses of the neutrinos of 
order $0.1$ eV, we need the ratio: $\frac{\mu}{M_X} \approx 10^{-4} $ assuming that $f \sim 10^{-4}$. The smallness 
of $\mu$ in comparison to the mass scale of heavy scalar doublets justifies the soft breaking of $Z_2$ symmetry. 

\section{Observed baryon asymmetry and DM abundance}\label{Obs_asy_and_DM_abun}
%%%%%%%%%%%%%%%%%%%%%%%%%%%%%%%%%%%%%%%%%%%%%%%%%%%%%%
The observed baryon asymmetry of the present Universe, usually reported in terms of the ratio of baryon to photon 
number density, $\eta \equiv n_B/n_\gamma$, is given as~\cite{pdg}
\begin{equation}\label{B-asy}
5.8\times 10^{-10} \leq \eta \leq 6.6\times 10^{-10} (\rm BBN) ~~~(95\% C.L)
\end{equation}
where $\eta = 7.04 Y_B$ with $Y_B=n_B/n_s$.
The ratio of DM to baryon abundance measured by WMAP and PLANCK in the cosmic microwave background is given to be 
$\frac{\Omega_{\rm DM}}{\Omega_B}\approx 5$, where $\Omega_i = \rho_i/\rho_c$, and $\rho_c$ is the critical density of the Universe. 
Thus the DM to baryon ratio can be rewritten as: 
\begin{equation}\label{DM_baryon_ratio}
\frac{\Omega_{\rm DM}}{\Omega_B}=\left(\frac{m_{\rm DM}}{m_p} \right) \left( \frac{Y_{\rm DM}}{Y_B}\right)
\end{equation}
where $m_{\rm DM}$ and $m_p$ are respectively DM and proton mass, $Y_{\rm DM}=n_{\rm DM}/s$ and $Y_B=n_B/s$ are 
respectively DM and baryon abundance in a comoving volume. In our case, the total DM abundance is sum of 
asymmetric and symmetric components as the DM remains out-of-equilibrium through out the epochs. Therefore, Eq. \ref{DM_baryon_ratio} 
can be rewritten as: 
\begin{equation}\label{DM_baryon_ratio_1}
\frac{\Omega_{\rm DM}}{\Omega_B}=\left(\frac{m_{\rm DM}}{m_p} \right) \left( \frac{Y_{\rm DM}^{\rm sym}}{Y_B} + \frac{Y_{\rm DM}^{\rm asy}}{Y_B}\right)\,.
\end{equation} 
As we discuss below the baryon asymmetry and DM abundance are resulted from the decay of a heavy scalar $X$, which we assume to 
be present in the early Universe. Therefore, the symmetric and asymmetric component of DM abundance as well as baryon asymmetry 
from X-decay can be approximately computed as: 
\begin{eqnarray}
Y_{DM}^{\rm asy} &=& Y_X \epsilon_\chi B_\chi \nonumber\\
Y_{DM}^{\rm sym} &=& Y_X B_\chi \nonumber\\
Y_B &=& c Y_L = c Y_X \epsilon_L B_L 
\label{DM&B-abundance}
\end{eqnarray}
where $c=-0.55$ is the fraction of lepton asymmetry that is converted to a net baryon asymmetry, $Y_X=n_X/s$ is the number density of $X$ 
in a comoving volume, $\epsilon_\chi$, $\epsilon_L$ are the CP-asymmetry parameters resulted through the 
decay of $X$ to $\psi\chi$ and $\nu_R \ell$ respectively, $B_\chi$, $B_L$ are the branching fractions for the decay of X to $\psi\chi$ and 
$\nu_R \ell$ respectively. Using Eq. \ref{DM&B-abundance} in \ref{DM_baryon_ratio_1} we get the DM to baryon ratio: 
\begin{equation}\label{DM_baryon_ratio_2}
\frac{\Omega_{\rm DM}}{\Omega_B}=\left(\frac{m_{\rm DM}}{m_p} \right) \left( \frac{B_\chi}{c \epsilon_L B_L} + \frac{B_\chi \epsilon_\chi}{c \epsilon_L B_L}\right)\,.
\end{equation}
The branching fractions $B_L$, $B_\chi$ and the CP-asymmetry parameters $\epsilon_L$, $\epsilon_\chi$ satisfy the 
constraints: 
\begin{equation}\label{constraints}
B_L+B_\chi = 1\,,~~~~\epsilon_L =-\epsilon_\chi=\epsilon ~~~~~~{\rm and} ~~~~~~~\epsilon_i \leq 2 B_i
\end{equation}
where the first constraint simply demands the unitarity of the model, while second and third constraints ensures that all 
the amplitudes are physical and total amount of CP-violation can not exceed 100\% in each channel. Using the above constraints 
in Eq. \ref{DM_baryon_ratio_2} we get 
\begin{equation}\label{DM_baryon_ratio_3}
\frac{\Omega_{\rm DM}}{\Omega_B}=\left(\frac{m_{\rm DM}}{m_p} \right) \frac{B_\chi}{c B_L} \left( \frac{1}{\epsilon_L } +1 \right)\,,
\end{equation}  
where the 1st term on the right-hand side is due to symmetric component while the second term comes from asymmetric component. For small 
Yukawa couplings (required for out-of-equilibrium condition of $X$) the CP asymmetry parameters, $\epsilon_i << 1$. This implies that the 
symmetric component always dominates over the asymmetric component unless in the resonance limit: $\epsilon_L \sim {\cal O}(1)$, where symmetric 
and asymmetric components contribute in similar magnitudes. Thus given the constraints \ref{constraints}, several comments are in order: 

(1) If $B_L=B_\chi=1/2$, then $M_\chi \approx {\cal O} (\epsilon) {\rm GeV}$. This implies from Eq. (\ref{DM&B-abundance}) that for
an optimistic case of $\epsilon=10^{-8}$, we get $M_\chi\approx 10$ eV.

(2) If $B_L >> B_\chi$ and $B_\chi \gtrsim \epsilon$ then we get $M_\chi \lsim 2.5$ GeV. 

(3) If $B_L << B_\chi$ then from  Eq. (\ref{DM&B-abundance}) we get  $10 {\rm eV} < M_\chi < 2.5$ GeV.  

Thus for a wide range of DM mass, we can generate correct relic abundance. In the following we solve the required Boltzmann equations 
to get the correct DM abundance, satisfying Eq. \ref{DM_baryon_ratio_3} by taking a typical DM mass to be 2.5 GeV, and 
observed baryon asymmetry given by Eq. \ref{B-asy}.    

%%%%%%%%%%%%%%%%%%%%%%%%%%%%%%%%%%%%%%%%%%%%%%%%%%%%%%%%%%%%%%%%%%%%%%%%%%%%%%
 \section{Lepton asymmetry from X-particles decay}\label{Production of LA}
%%%%%%%%%%%%%%%%%%%%%%%%%%%%%%%%%%%%%%%%%%%%%%%%%%%%%%%%%%%%%%%%%%%%%%%%%%%%%%
We assume that the $X$-particle is present in the early Universe. At a temperature above its mass 
scale, $X$ is in thermal equilibrium due to gauge and Yukawa mediated interactions. As the temperature falls, due 
to Hubble expansion, below the mass scale of X-particle, the latter goes out-of-thermal equilibrium and decay. The 
decay rate of $X$-particle can be given as: 
\begin{equation}
\Gamma_X \simeq \frac{1}{8\pi}(f^2 + \lambda^2) M_X \,,
\end{equation}
 where $f$, $\lambda$ are Yukawa couplings. Demanding that $\Gamma_X \lesssim H$ at $T=M_X$, 
where $H=1.67 g_*^{1/2} T^2/M_{\rm Pl}$ is the Hubble scale of expansion, we get $M_X \lesssim 10^{10}$ GeV for $f \sim \lambda 
\lsim 10^{-4}$. The CP violating decay of $X$ requires at least two copies. More over, we assume a mass hierarchy between the two X-particles 
so that the CP-violating decay of lightest $X$ to $\ell \nu_R$ and $\psi \chi$ generates asymmetries both in lepton and 
dark matter sectors. Thus the decay modes to visible sector are $X^0\to \nu_L \nu_R$ and $X^- \to \ell^- \nu_R$, while the 
decay modes to dark matter sector are $X^0 \to \psi^0 \chi$ and $X^- \to \psi^- \chi$. The two decay modes of $X$ to 
visible and dark sectors are equivalent and hence we will focus only one of the channels, say $X^- \to \ell^- \nu_R$ in 
the visible sector and $X^- \to \psi^- \chi$ to dark sector. 
\begin{figure} [h!]
\centering
\includegraphics[width=80mm]{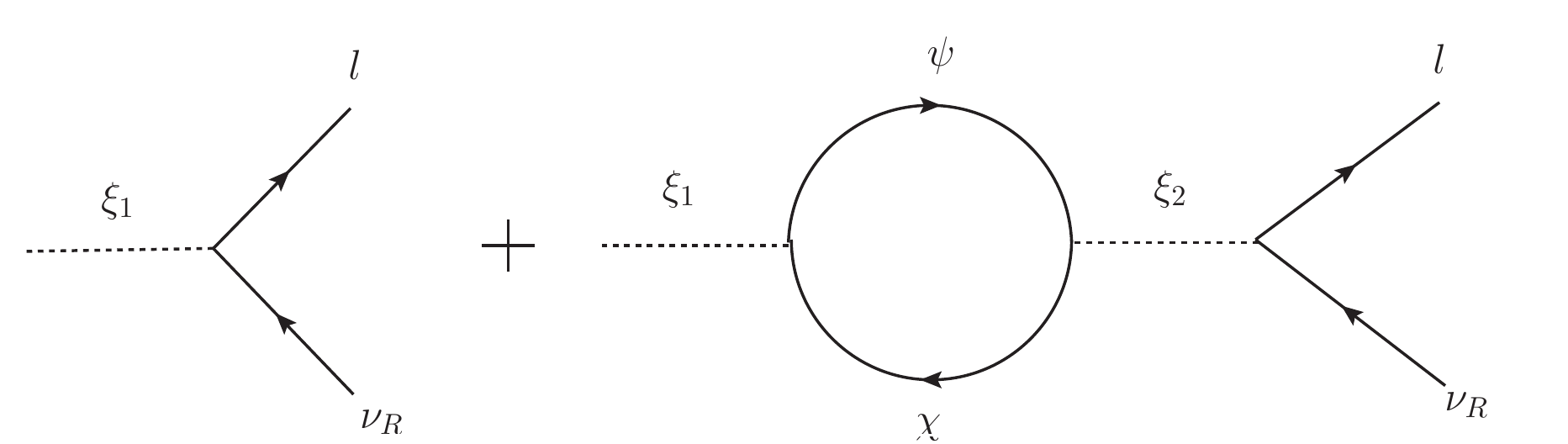}
\caption{\footnotesize{Tree level and self energy correction diagrams, whose interference give rise to a net CP violation.}}
\label{LA}
\end{figure}
In presence of $X$-particles and their interactions, the diagonal mass terms in Eq.~\ref{potential} can be replaced by~\cite{Ma:1998dx,Arina:2011cu},
\begin{equation}
\frac{1}{2} X^{\dagger}_{a}(M_{+}^{2})_{ab}X_{b}+\frac{1}{2} (X^{*}_{a})^{\dagger}(M_{-}^{2})_{ab}X^{*}_{b}\,,
\end{equation}
Where
\begin{equation}
M^{2}_{\pm}= 
\begin{pmatrix} M_{1}^{2}-i C_{11} & -i C_{12}^{\pm}\cr\\
-i C_{21}^{\pm} & M_{2}^{2}-i C_{22} \end{pmatrix}
\label{mass_matrix}
\end{equation}
here
$C_{ab}^{+}=\Gamma_{ab}M_{b}$, $C_{ab}^{-}=\Gamma^{*}_{ab}M_{b}$ and $C_{aa}=\Gamma_{aa}M_{a}$ with
\begin{equation}
\Gamma_{ab}M_{b}=\frac{1}{8\pi}\left(M_{a}M_{b}\lambda_{a}\lambda_{b}^{*}+M_{a}M_{b} \mathop{\sum_{k,l}} f_{akl}^{*}f_{bkl}\right)
\end{equation}

Diagonalizing the above mass matrix Eq.~\ref{mass_matrix}, we get two mass eigenvalues $M_{\xi_1}$ and $M_{\xi_2}$ corresponding to the two 
eigenstates $\xi_1^{\pm}$ and $\xi_2^{\pm}$. Note that the mass eigenstates $\xi_1^+$ and $\xi_1^-$ (similarly $\xi_2^+$ and $\xi_2^-$) 
are not CP conjugate states of each other even though they are degenerate mass eigen states. Hence their decay can give rise 
to CP-asymmetry. The CP-violation arises via the interference of tree level and one loop self energy correction diagrams shown in 
Fig.~\ref{LA}. The asymmetry in the visible sector is given by
\begin{eqnarray}
\epsilon_{L} &=& [B_L(\xi_{1}^{-} \rightarrow l^{-} \nu_{R})-B_L(\xi_{1}^{+}\rightarrow (l^{-})^{c} \nu_{R}^{c})]\nonumber\\
  &=& -\frac{Im\left(\lambda_{1}^{*} \lambda_{2} \mathop{\sum_{k,l}} f^{*}_{1kl} f_{2kl} \right)}{8\pi^2(M_{2}^{2}-M_{1}^{2})} \left[\frac{M_{1}^{2}M_{2}}{\Gamma_{1}} \right]\,,
 \end{eqnarray}
where $B_L$ is the branching ratio for $\xi_{1}^{\pm} \rightarrow l^{\pm} \nu_{R}$. Using the CP-asymmetry $\epsilon_{L}$, we can estimate the generated 
lepton asymmetry $Y_{L} \equiv \frac{n_{L}}{s}$, where $s=(2\pi^2/45)g_* T^3$ is the entropy density, from the decay of $\xi_1$. The relevant Boltzmann equations governing the 
evolution of the number density of $\xi_1$, {\it i.e.} $Y_{\xi_{1}}$, and lepton asymmetry $Y_L$ are given by:

\begin{equation}
\frac{dY_{\xi_{1}}}{dx}=-\frac{x}{H(M_{\xi_{1}})}s<\sigma|v|_{(\xi_{1}\xi_{1}\rightarrow All)}>\left[ Y_{\xi_{1}}^{2}-{Y_{\xi_{1}}^{eq}}^2 \right]-\frac{x}{H(M_{\xi_{1}})}\Gamma_{(\xi_{1} \rightarrow All)} \left[ Y_{\xi_{1}}-Y_{\xi_{1}}^{eq} \right]
\end{equation}
 and 
\begin{equation}
\frac{dY_{L}}{dx}=\epsilon_{L} \frac{x}{H(M_{\xi_{1}})}\Gamma_{(\xi_{1}\rightarrow All)}B_{L} \left[ Y_{\xi_{1}}-Y_{\xi_{1}}^{eq} \right]\,,
\end{equation}
where the $x=\frac{M_{\xi_{1}}}{T}$, is the dimensionless variable which ranges from $ 0 \to \infty$ as the temperature $T: \infty \to 0$. 
\begin{figure} [h!]
\centering
\includegraphics[width=90mm]{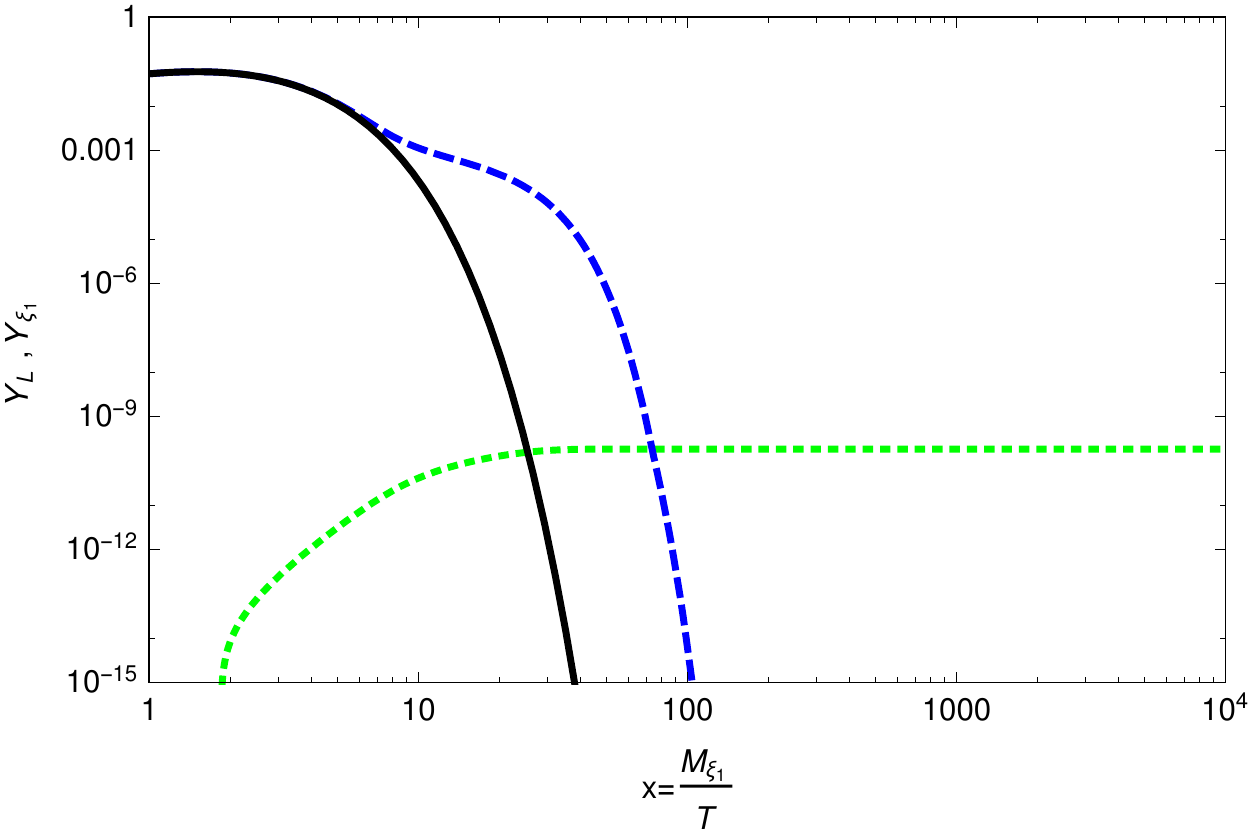}
\caption{\footnotesize{The lepton asymmetry from the decay of $\xi_1$. The Green(dotted) line shows the abundance of lepton asymmetry 
for $\epsilon_{L} = 2 \times 10^{-7}$. The Blue(dashed) line shows the abundance of $\xi_1$ particles. The Black solid line shows the 
equilibrium number density of $\xi_1$. }}
\label{yL_yDM}
\end{figure}
We have shown in Fig.~\ref{yL_yDM} the lepton asymmetry $Y_{L}$ and the comoving number density of $\xi_{1}$, {\it i.e.} $Y_{\xi_{1}}$, 
as function of $x$. The decay coupling constant $f$ is taken as $10^{-4}$ and the $\lambda$ is taken as $0.47 \times 10^{-7}$. The typical value of 
the cross-section is taken as $\sigma|v|_{(\xi_{1}\xi_{1} \rightarrow All)}=10^{-25} GeV^{-2}$. The blue (dashed) line shows the abundance of 
$\xi_{1}$ particles. From Fig. ~\ref{yL_yDM}, we see that as the temperature falls below the mass of $\xi_1$ ({\it i.e. $x > 1$}), it decouples 
from the thermal bath and then decays. The lepton asymmetry, which is proportional to $B_L$ and $\epsilon_L$, starts developing as $\xi_{1}$ 
decays to $\ell^- \nu_R$ and settles to a constant value after the decay of $\xi_1$ is completed. In Fig.~\ref{yL_yDM} we have taken the branching 
ratio $B_L \sim {\cal O}(1)$ and the heavy scalar mass, $M_{\xi_{1}}$ = $10^{10}$ GeV. The electroweak sphalerons which violates $B+L$, but 
conserves $B-L$, can transfer a partial lepton asymmetry to a net baryon asymmetry $Y_{B} = -0.55 Y_{L}$. For the $\epsilon_{L}$ = $ 2 \times 10^{-7}$ 
and the parameters we discussed above, we get the baryon asymmetry, $Y_{B} = -1 \times 10^{-10}$.

%%%%%%%%%%%%%%%%%%%%%%%%%%%%%%%%%%%%%%%%%%%%%%%%%%%%%%%%%%%%%%%%%%%%%%%%%%%%
\section{Dark matter abundance from X-particles decay}\label{Dark matter}
%%%%%%%%%%%%%%%%%%%%%%%%%%%%%%%%%%%%%%%%%%%%%%%%%%%%%%%%%%%%%%%%%%%%%%%%%%%%%
The decay of $X$-particles to $\psi \chi$ can populate the number densities of $\psi$ and $\chi$. Since $\psi$ is a doublet 
under $SU(2)_L$, it thermalises quickly via gauge interactions. As we will discuss in section \ref{sym_abun}, the symmetric 
component of $\psi$ gets annihilated efficiently to the SM particles, while the asymmetric number density of $\psi$ 
gets converted to a net $\chi$ density through the decay process: $\psi \to \chi \bar{f} f$, induced via the soft 
$Z_2$ symmetry breaking term $\mu^2 H^\dagger X$. Here we assume $M_{\chi} < M_{\psi}$, so that $\chi$ remains stable 
kinematically. On the other hand, $\chi$ is a singlet under $SU(2)_L$. The only way 
$\chi$ can interact with the SM particles is via the mixing of neutral gauge bosons $Z_D$ and $Z$. However, as we show later 
that $\chi$ remains out-of-equilibrium through out the epoch. As a result the symmetric component of $\chi$ does not get 
annihilated efficiently to the SM particles like $\psi$. Therefore, the number density of the symmetric component of $\chi$ 
always dominates over its asymmetric number density. In the following we discuss the symmetric and asymmetric abundance 
of $\chi$ produced via the decay of $X$-particles.

%%%%%%%%%%%%%%%%%%%%%%%%%%%%%%%%%%%%%%%%%%%%%%%%%%%%%%%%%%%%%%%%%%%%%%%%%%%%%%%%%%%%%%%%%%%%%%%%%%%
\subsection{Symmetric $\chi$-DM abundance from X-particles decay}\label{sym_abun}
%%%%%%%%%%%%%%%%%%%%%%%%%%%%%%%%%%%%%%%%%%%%%%%%%%%%%%%%%%%%%%%%%%%%%%%%%%%%%%%%%%%%%%%%%%%%%%%%%%%
Let us now discuss about the symmetric component of the $\chi$-abundance. Note that $\chi$ is a singlet under electroweak 
interaction. Therefore, we safely assume that the thermal abundance is negligible. In this case, the number density of 
$\chi$ particles is produced by the CP-conserving decay of heavy scalar $\xi_{1}$. In the early Universe, when $\xi_{1}$ 
goes out of thermal equilibrium and decay to $\psi$ and $\chi$, the $\psi$ gets thermalised quickly through its gauge 
interaction, while the $\chi$ remain isolated. The abundance of $\chi$ from decay of $\xi_1$ and $\xi_2$ can be estimated 
from the following Boltzmann equations:  
\begin{equation}\label{eqn.1}
\frac{dY_{\xi_{1}}}{dx}=-\frac{x}{H(M_{\xi_{1}})}s<\sigma|v|_{(\xi_{1}\xi_{1} \rightarrow All)}>\left[ Y_{\xi_{1}}^{2}-{Y_{\xi_{1}}^{eq}}^2 \right]-\frac{x}{H(M_{\xi_{1}})}\Gamma_{(\xi_{1} \rightarrow All)} \left[ Y_{\xi_{1}}-Y_{\xi_{1}}^{eq} \right]\,,
\end{equation}
and 
\begin{equation}\label{eqn.2}
\frac{dY_{\chi}}{dx}=\frac{x}{H(M_{\xi_{1}})}\Gamma_{(\xi_{1}\rightarrow All)} B_{\chi} \left[ Y_{\xi_{1}}-Y_{\xi_{1}}^{eq} \right].
\end{equation}
%%%%%%%%%%%%%%%%%%%%%%%%%%%%%%%%%%%%%%%%%%%%%%%%%
\begin{figure} [h!]
\centering 
\includegraphics[width=90mm]{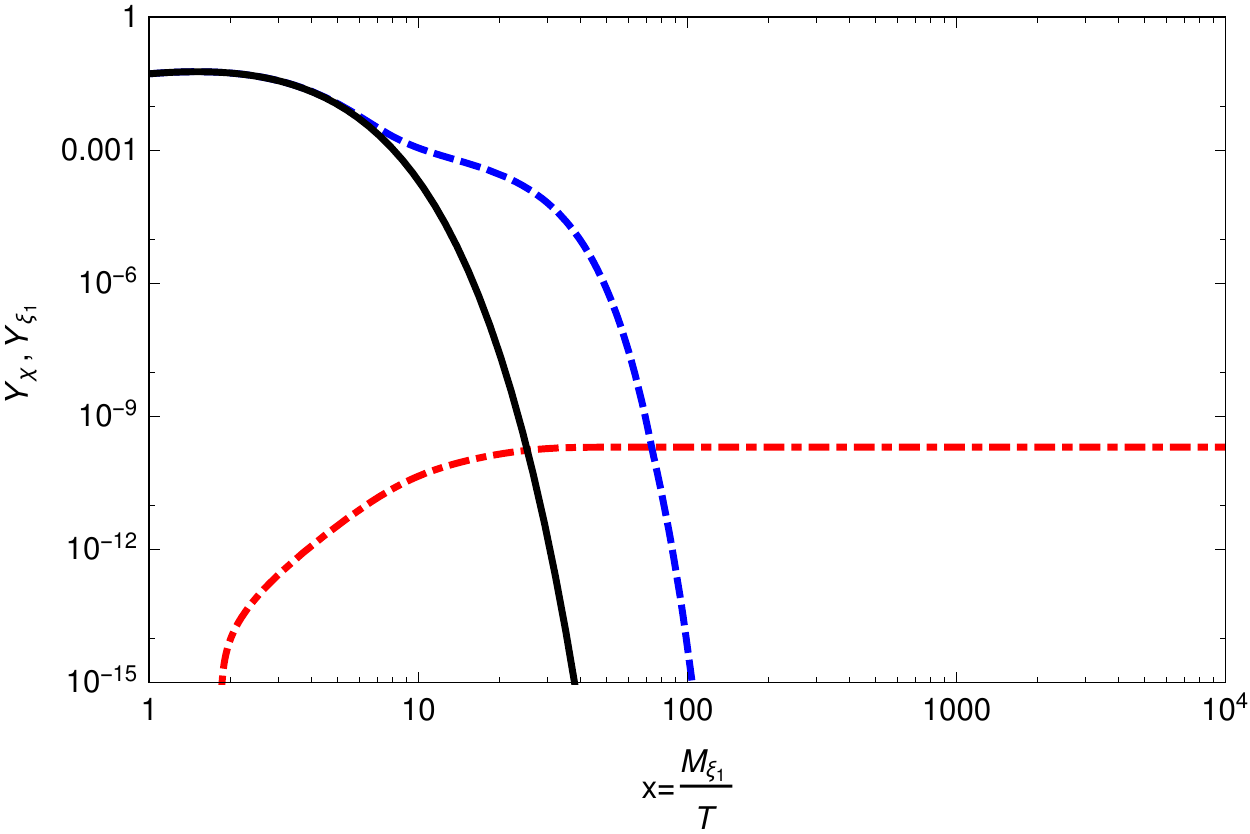}
\caption{\footnotesize{The Red dotted-line shows the abundance of $\chi$ dark matter and the Blue dashed-line shows the abundance of $\xi_1$ particles. 
The Black solid-line shows the equilibrium number density of $\xi_1$.}}
\label{Dark_matter_abundance}
\end{figure}
%%%%%%%%%%%%%%%%%%%%%%%%%%%%%%%%%%%%%%%%%%%%%%%%%
In the above Eq.~\ref{eqn.2} , $B_\chi \equiv Br(\xi_1\to \psi \chi)$ is the branching ratio for the decay of $\xi_1$ to $\psi \chi$. The solutions of Eqs. \ref{eqn.1} and \ref{eqn.2} are 
shown in Fig.~\ref{Dark_matter_abundance} . Here we use the decay coupling constant $\lambda$ = $0.47 \times 10^{-7}$, $M_{\xi_{1}}$ = $10^{10}$ GeV 
and $M_{\chi}$ = 2.5 GeV. The typical value of the cross section for $\xi_1^\dagger \xi_1 \to {\rm All particles}$ is taken as 
$\sigma|v|_{(\xi_{1}\xi_{1} \rightarrow All)}=10^{-25}  GeV^{-2}$. The blue dashed-line shows the abundance of $\xi_{1}$ particles, where 
as the red dot-dashed line shows the abundance of dark matter particle $\chi$. To get the observed dark matter abundance,
\begin{equation}
Y_{DM} \equiv \frac{n_{DM}}{s} = 4 \times 10^{-12} \left(\frac{100 GeV}{M_{DM}} \right)\left(\frac{\Omega_{DM} h^{2}}{0.11} \right)
\end{equation}
we have used the branching ratio: $B_\chi = 2.2 \times 10^{-7}$. This shows that $\xi_1$ decay significantly 
to leptons and rarely to invisible sector to get the correct relic abundance of dark matter and baryon asymmetry.

As the temperature falls below the mass scale of $\psi$, the latter gets decoupled from the thermal bath and decay back to $\chi$ 
and may produce an additional abundance of dark matter. However, as we show below the freeze-out cross-section of $\psi \overline{\psi} 
\to {\rm SM particles}$ is quite large due to its coupling with SM gauge bosons and hence produce a significantly low abundance. The 
relevant Boltzmann equation for the evolution of $\psi$ number density is given by:

\begin{equation}
\frac{dY_{\psi}}{dz}=-\frac{z s}{H(M_{\psi})}<\sigma|v|>_{\rm eff} \left[ Y_{\psi}^{2}-{Y_{\psi}^{eq}}^{2} \right]\,,
\end{equation}

where the $z=\frac{M_{\psi}}{T}$, and s is the entropy density. The relevant channels contributing to the $\psi$ relic density are: 

$\overline{\psi^{+}}\psi^{+} \rightarrow \gamma Z_{d}, \gamma \gamma, W^{+} W^{-} \bar{u} u, \bar{c} c, \bar{t} t, \bar{l} l$ ,

 $\psi^{+} \overline{\psi^{0}} \rightarrow W^{+} Z_{d}, \bar{b} t, \bar{d} u , \bar{s} c , \gamma W^{+}, \bar{l} \nu_{l}, Z W^{+}$ and

 $\overline{\psi^{0}}\psi^{0} \rightarrow Z Z_{d}, Z Z, W^{+} W^{-}, \bar{q} q$ .

\begin{figure} [h!]
\centering
\includegraphics[width=80mm]{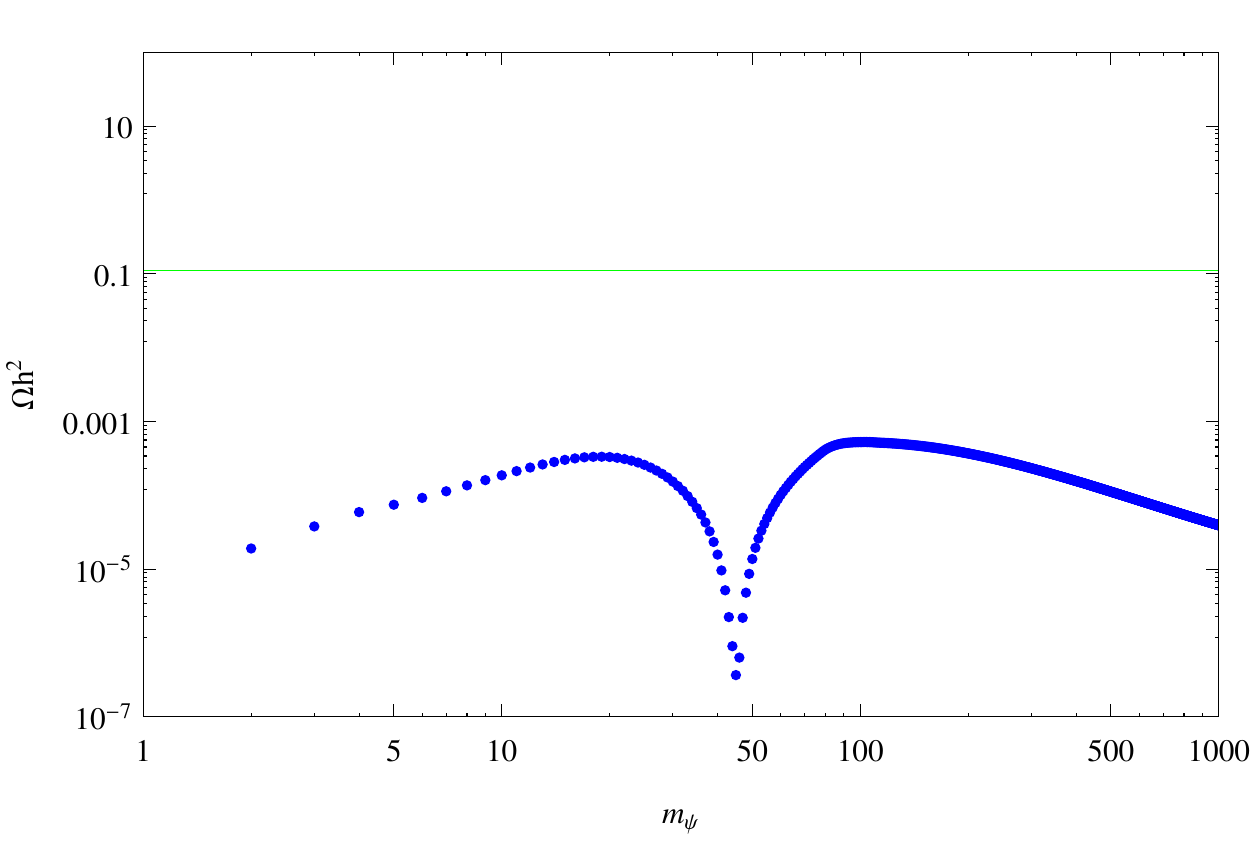}
\caption{\footnotesize{Relic abundance of $\psi$ particles (shown by dotted Blue line). Green horizontal line shows the observed relic abundance by PLANCK data.}}
\label{relicdpsi}
\end{figure}

We use micrOMEGAs~\cite{Belanger:2008sj} to calculate the freeze-out abundance of $\psi$ particles. The results are shown in 
Fig.~\ref{relicdpsi} of ref. ~\cite{Bhattacharya:2017sml}.

One can clearly see in Fig.~\ref{relicdpsi}, that the resonance at $M_{\psi}$ = $\frac{M_{Z}}{2}$, is due to the enhancement in the Z mediation s-channel 
cross section, which causes a drop in the relic density. More over we see that the relic density of $\psi$ is much less than the observed DM abundance 
by WMAP~\cite{Hinshaw:2012aka} and PLANCK~\cite{Ade:2015xua}. Therefore, the decay of $\psi$ after it freezes out does not produce any 
significant $\chi$ abundance.

\subsection{Asymmetric $\chi$-DM abundance from X-particles decay}
\begin{figure} [h!]
\centering
\includegraphics[width=80mm]{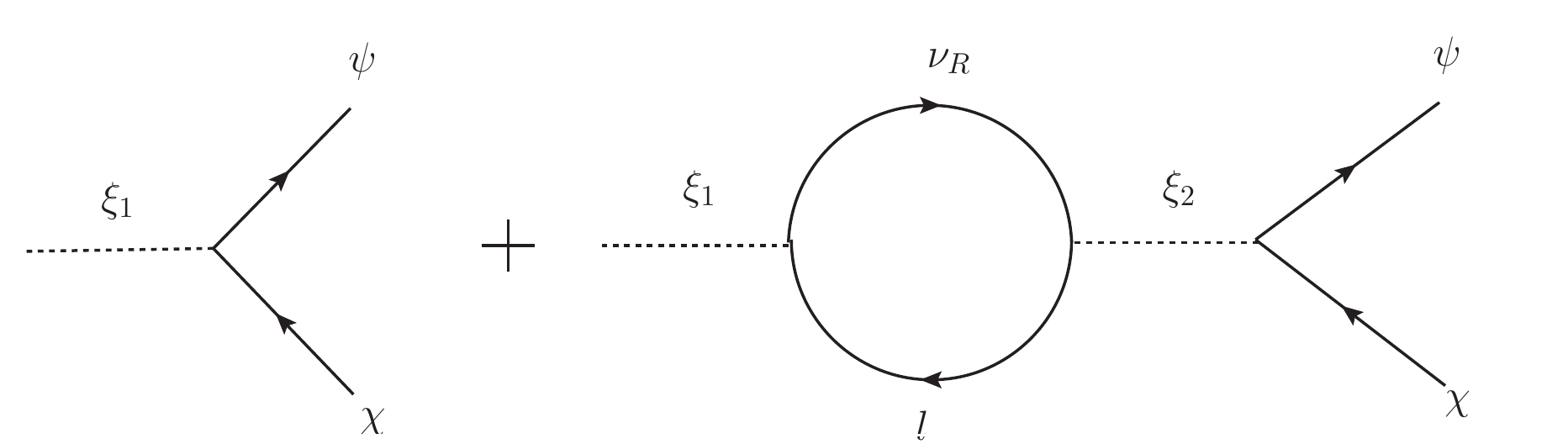}
\caption{\footnotesize{Tree level and self energy correction diagrams producing the dark matter asymmetry}}
\label{DA}
\end{figure}

%%%%%%%%%%%%%%%%%%%%%%%%%%%%%%%%%%%%%%%%%%%%%%
Similar to the lepton asymmetry, the CP-violating out-of-equilibrium decay of $\xi_1$ produce an asymmetry between 
$\chi$ and $\bar{\chi}$ as well as $\psi$ and $\bar{\psi}$. The corresponding tree level and self energy correction 
diagrams are shown in Fig. \ref{DA}. 

The amount of CP-asymmetry can be given as:
\begin{eqnarray}
\epsilon_\chi &=& [Br(\xi_{1}^{-} \rightarrow \psi^{-} \chi)-Br(\xi_{1}^{+}\rightarrow (\psi^{-})^{c} \chi^{c})]\nonumber\\
    &=& \frac{Im\left(\lambda_{1}^{*} \lambda_{2} \mathop{\sum_{k,l}} f^{*}_{1kl} f_{2kl} \right)}{8\pi^2(M_{2}^{2}-M_{1}^{2})} \left[\frac{M_{1}^{2}M_{2}}{\Gamma_{1}} \right]=-\epsilon_L.
 \label{DA_asy2}
 \end{eqnarray}
The corresponding $\chi$ asymmetry can be computed as: 
\begin{equation}
Y_\chi= \epsilon_\chi Y_{\xi_1} B_\chi\,.
\end{equation}
Since $|\epsilon_L|=|\epsilon_\chi|= 10^{-7}$ (required for observed baryon asymmetry; see section \ref{Production of LA}) 
and $B_\chi={\cal O} (10^{-7})$ (required for observed DM abundance; See section \ref{sym_abun}) we get a very small asymmetry, 
${\cal O}(10^{-16})$. Moreover, $\psi$ and $\chi$ are vector-like fermions. Therefore, the sphalerons don't convert this asymmetry 
to a net baryon asymmetry. See for instance~\cite{Arina:2011cu,Arina:2012fb,Arina:2012aj}. Thus the observed baryon asymmetry does 
not get affected by the decay of $\xi_1$ to $\psi \chi$.   

\subsection{Production of $\chi$-DM from thermal scattering}
%%%%%%%%%%%%%%%%%%%%%%%%%%%%%%%%%%%%%%%%%%%%%%%%%%%%%%%%%%%%%%%%%%
\begin{figure} [h!]
\centering
\includegraphics[width=60mm]{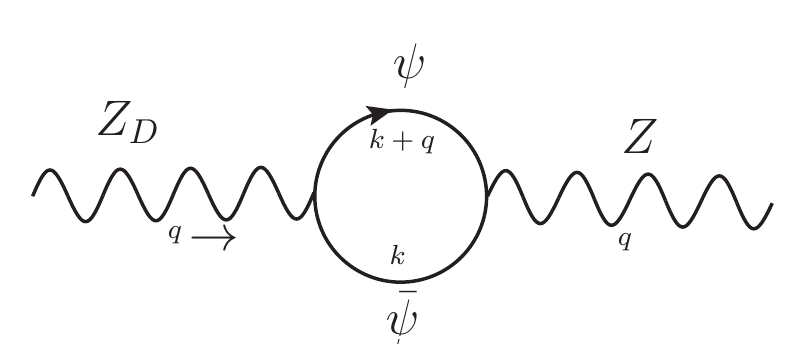}
\caption{\footnotesize{The mixing between the Z and $Z_{D}$ with the running of $\psi$} particles in the loop.}
\label{zpsizmixing}
\end{figure}

Now we check the possible scattering processes through which $\chi$ can be produced in the thermal bath apart from $X$-decay.
The relevant processes are $\overline{\psi} \psi \to \overline{\chi}\chi$ mediated via $Z_D$ and $f \bar{f} \to \overline{\chi}\chi$ via 
the exchange of $Z-Z_D$ mixing as shown in the Fig.~\ref{zpsizmixing}. The former one is the most relevant one as it dominates over the 
latter process. The scattering cross-section times velocity for the process: $\overline{\psi}\psi \to \overline{\chi}\chi$ is given by:  
\begin{equation}
\sigma = \frac{ \sqrt{s-4 M_{\chi}^{2} }}{16 \pi s \sqrt{s}} \frac{ g_{D}^{4} }{ (s-M_{Z_{D}}^{2})^{2}+\Gamma_{Z_{D}}^{2}M_{Z_{D}}^{2} }
\left( s^2 + \frac{1}{3}(s-4 M_{\psi}^{2})(s-4 M_{\chi}^{2}) + 4 M_{\chi}^{2} s + 4 M_{\psi}^{2} s  \right)\,.
\label{psipsi-to-chichi}
\end{equation}
In the  process: $f \bar{f} \to \overline{\chi}\chi$ via the exchange of $Z-Z_D$ mixing, the loop factor is estimated 
as~\cite{Patra:2016shz}:
\begin{equation}
\Pi^{\mu \nu}(q) = \left(q^{2} g^{\mu \nu}-q^{\mu} q^{\nu} \right)\frac{g_{D}}{4\pi^{2}}\left(\frac{g}{2}\cos\theta_{W}-\frac{g'}{2} \sin\theta_{W}  \right)\int_{0}^{1}dx 2x(1-x)\log\left( \frac{M_{\psi}^{2}}{M_{\psi}^{2}-x(1-x)q^{2}}\right).
\end{equation}
Where $\theta_{W}$ is the Weinberg angle, the $M_{\psi}$ is the mass of the $\psi$ particle running in the loop. Now the cross-section times velocity 
for this process is given as:
\begin{eqnarray}
\sigma|v| &=& \frac{\sqrt{s-4M_{\chi}^{2}}}{2 \pi s \sqrt{s}}  \frac{\left(\frac{g}{2\cos{\theta_{W}}}g_{D}\hat{\Pi_{2}}(s)\right)^{2}}{\left[(s-M_{Z}^{2})^{2}+\Gamma_{Z}^{2}M_{Z}^{2}\right] \left[(s-M_{Z_{D}}^{2})^{2}+\Gamma_{Z_{D}}^{2} M_{Z_{D}}^{2} \right]} \nonumber\\
&& \left[\left(g_{V}^{2}+g_{A}^{2}\right)\left\lbrace\frac{5}{12}s^{4}-\frac{1}{3}M_{\chi}^{2}s^{3}-\frac{2}{3}M_{f}^{2}s^{3}+\frac{2}{3}M_{f}
^{2}M_{\chi}^{2}s^{2}\right\rbrace+ \left(g_{V}^{2}-g_{A}^{2}\right) \left\lbrace M_{f}^{2}s^{3}+2M_{f}^{2}M_{\chi}^{2}s^{2}\right\rbrace \right]\,.
\label{sigma_v}
\end{eqnarray}
In the limit $s>4M_{\psi}^{2}$,
\begin{equation}
\hat{\Pi_{2}}(s) = \frac{4g_{D}(\frac{g}{2}\cos\theta_{W}-\frac{g'}{2}\sin\theta_{W})}{16 \pi^{2}}(\pm \pi)\frac{1}{6}\sqrt{1-\frac{4M_{\psi}^{2}}{s}} \left(1+\frac{2M_{\psi}^{2}}{s}\right)\,.
\end{equation}
If the above mentioned processes are brought to thermal equilibrium then they will overpopulate the $\chi$-number density. Therefore, we need to 
check the parameter space in which the above processes remain out-of-equilibrium through out the epochs. Note that in 
Eqs.~\ref{psipsi-to-chichi} and \ref{sigma_v}, the unknown parameters are $g_D$ and $M_{Z_D}$ apart from $M_\chi$. In Fig.~\ref{gdmzd}, we have 
shown the parameter space (given by Blue region) in the plane of $g_{D}$ versus $M_{Z_{D}}$, where the above processes remain out-of-equilibrium 
and hence remain consistent with the dark matter relic abundance obtained from $X$-decay. In this case, we have chosen a typical center of mass 
energy: $\sqrt{s}=1000$ GeV. The sharp deep at around $M_{Z_D}=1000$ GeV, implies that the cross-section is large at the resonance.

\section{Direct search of $\chi$ dark matter}\label{DD}
\begin{figure} [h!]
\centering
\includegraphics[width=50mm]{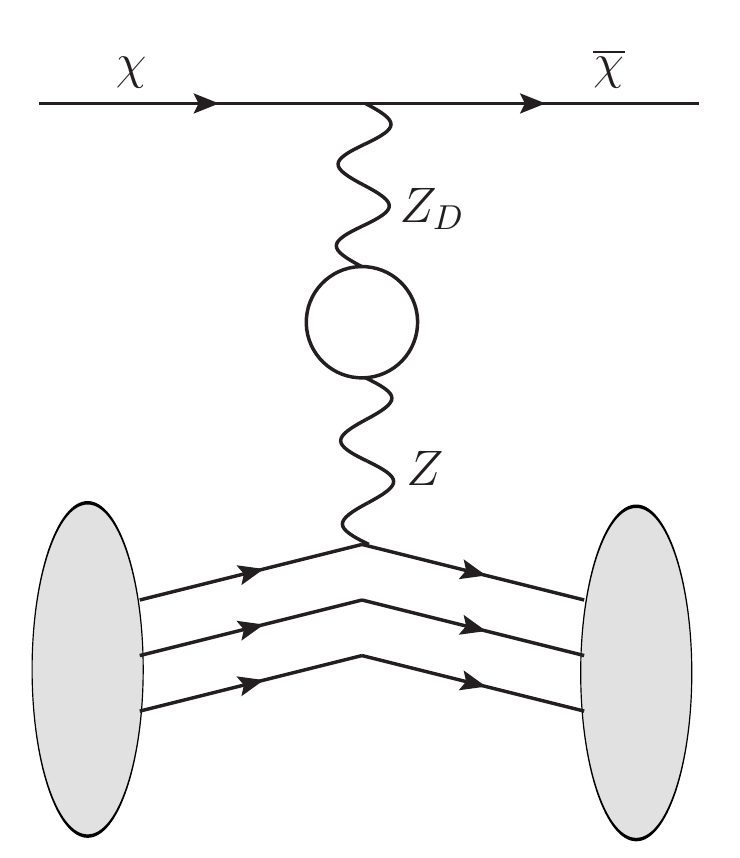}
\caption{\footnotesize{Dark matter scattering with nuclei via the $Z-Z_{D}$ mixing}.}
\label{Direct Detection}
\end{figure}

%%%%%%%%%%%%%%%%%
The spin independent elastic cross-section of DM candidate with nuclei through the $Z-Z_{D}$ mixing is shown in the Fig.~\ref{Direct Detection}.

\begin{figure} [h!]
\centering
\includegraphics[width=80mm]{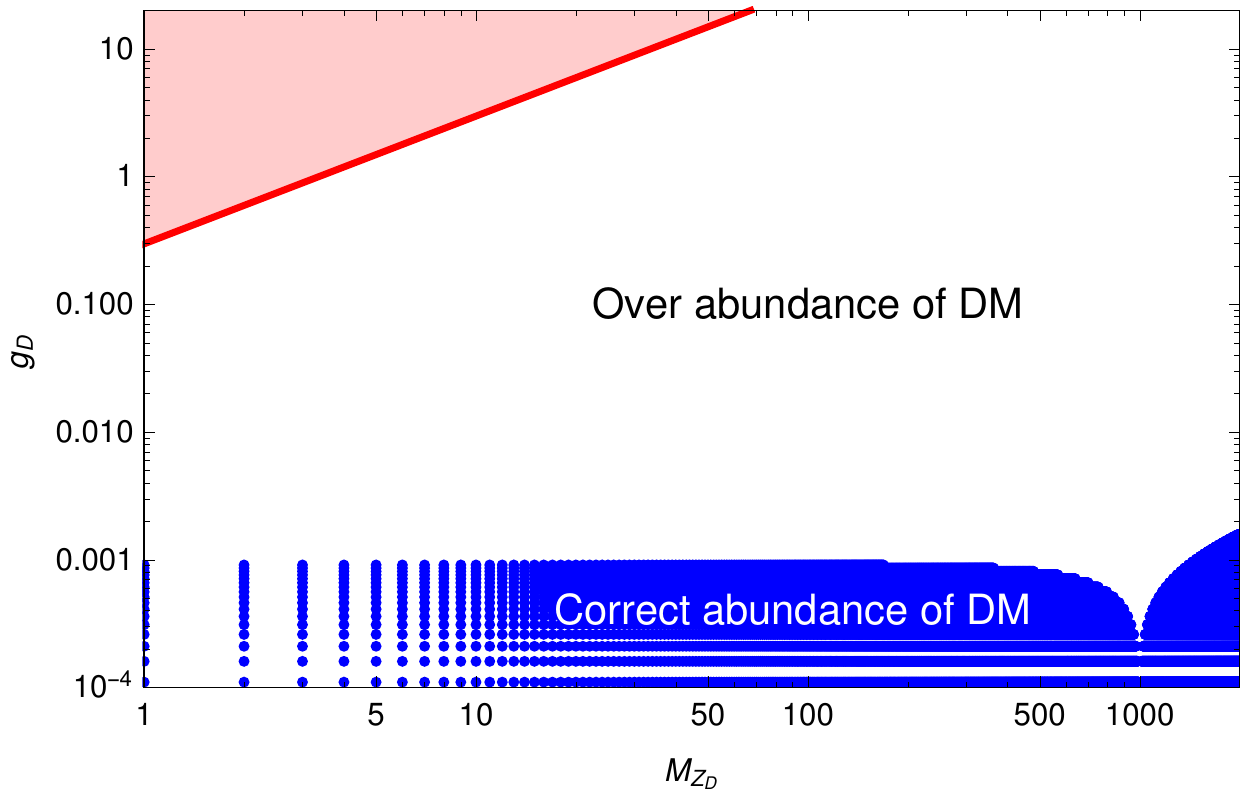}
\caption{\footnotesize{LUX constraint on dark matter, arising via $Z-Z_{D}$ mixing, is shown on the plane of $g_{D}$ versus $M_{Z_{D}}$ 
for a typical dark matter mass of 5 GeV(top Red line), using $\tan\theta_{Z}=10^{-2}$. The blue region defines the allowed parameter space 
which is consistent with the dark matter relic abundance.}}
\label{gdmzd}
\end{figure}

The spin independent DM-nucleon cross-section with loop induced $Z-Z_{D}$ mixing is given by~\cite{Goodman:1984dc}~\cite{Essig:2007az},
 \begin{equation}
\sigma_{SI}^Z = \frac{1}{64 \pi A^{2}} \mu_{r}^{2} \tan^{2}{\theta_{Z}} \frac{G_{F}}{2\sqrt{2}} \frac{g_{D}^{2}}{M_{Z_{D}}^{2}}\left[\tilde{Z} \frac{f_{p}}{f_{n}} +(A-\tilde{Z}) \right]^{2} f_{n}^{2}\,,
\label{sigmaeq}
\end{equation}
Where A is the mass number of the target nucleus, $\tilde{Z}$ is the atomic number of the target nucleus, $\theta_{Z}$ is the mixing angle 
between $Z$ and $Z_{D}$, $\mu_{r} = M_{\chi} m_{n}/(M_{\chi} + m_{n}) \approx m_{n}$ is the reduced mass, $m_{n}$ is the mass of nucleon 
(proton or neutron) and $f_{p}$ and $f_{n}$ are the interaction strengths of DM with proton and neutron respectively. For simplicity we 
assume conservation of isospin, i.e. $f_{p}/f_{n} = 1$. The value of $f_{n}$ is varied within the range: $0.14 < f_{n} < 0.66$~\cite{Koch}. 
If we take $f_{n} \simeq 1/3$, the central value, then from Eq.~\ref{sigmaeq} we get the total cross-section per nucleon to be,
\begin{equation}
\sigma_{SI}^{Z} \simeq 2.171 \times 10^{-36}cm^{2} \tan^{2}{\theta_{Z}} \frac{g_{D}^{2}}{ (M_{Z_{D}}/{\rm GeV})^{2}}\,,
\end{equation}
where we have used DM mass to be 5 GeV. Since the Z-boson mass puts a stringent constraint on the mixing parameter $\tan\theta_{Z}$ to 
be $\mathcal{O}(10^{-2}-10^{-4})$~\cite{Patra:2016shz,Hook:2010tw,Babu:1997st}, we choose the maximum allowed value $(10^{-2})$ and 
plot the spin independent direct DM detection cross-section, allowed by LUX \cite{Akerib:2016vxi}, in the plane of $g_{D}$ 
versus $M_{Z_{D}}$ as shown in Fig.~\ref{gdmzd}. The plot shows a straight line, as expected from Eq.~\ref{sigmaeq}, and is given 
by the Red lines in Fig.~\ref{gdmzd} for the DM mass of 5 GeV. Any values above that line corresponding to the DM mass of 5 GeV will 
not be allowed by the LUX limit. 

%%%%%%%%%%%%%%%%%%%%%%%%%%%%%%%%%%%%%%%%%%%%%%%%%%%%%%%%%%%%%%%%%%%%%%%%%%%%%%
\section{Conclusions}\label{Conclusion}
%%%%%%%%%%%%%%%%%%%%%%%%%%%%%%%%%%%%%%%%%%%%%%%%%%%%%%%%%%%%%%%%%%%%%%%%%%%%%%
The oscillation experiments undoubtedly shown that the neutrinos are massive. However, their nature, either Dirac or Majorana, 
is yet to be confirmed. In this paper, by assuming that the neutrinos are Dirac ($B-L$ is an exact symmetry), we found a way of 
explaining simultaneously the relic abundance of dark matter and baryon asymmetry of the Universe. 

We extended the SM with a simple dark sector constituting vector-like fermions: a doublet $\psi$ and a singlet $\chi$, where $\chi$ is 
odd under the discrete $Z_2$ symmetry and behave as a candidate of dark matter. The same $Z_2$ symmetry disallowed neutrino Dirac 
mass by forbidding $\bar{\nu_R} \tilde{H}^\dagger \ell$ coupling, where $\nu_R$ is odd under the $Z_2$ symmetry. However, the discrete 
$Z_2$ symmetry was allowed to break softly without destabilizing the dark matter component $\chi$ ({\it i.e.} we chose $M_\chi < M_\psi$). As a 
result, Dirac mass of the active neutrinos could be generated.  

We assumed heavy Higgs doublets ($X$), which transform non-trivially under the discrete $Z_2$ symmetry, present in the early Universe. 
The out-of-equilibrium decay of $X$ through $X \to \nu_R\ell$ and $X\to \chi \psi$ generated baryon asymmetry and dark matter abundance that we 
observe today. Since $B-L$ is considered to be an exact symmetry, the CP-violating decay of $X$ to $ \nu_R\ell$ produced equal and opposite 
$B-L$ asymmetries in the left and right-handed sectors. The right-handed sector coupled weakly to the SM as required by the Dirac mass of the 
neutrinos. Therefore, the $B-L$ asymmetry in the left-handed sector got converted to a net $B$-asymmetry via the $B+L$ violating sphaleron 
transitions, while that in the right-handed sector remained intact. The two $B-L$ asymmetries neutralized much after the electroweak phase 
transition when the sphaleron transitions got suppressed. Similarly the decay of $X \to \chi \psi$ also generated a net $\chi$ abundance that 
we observe today. Since the branching fraction $B_\chi << 1$, the asymmetric $\chi$ abundance is much smaller than its symmetric counterpart. 

The dark matter $\chi$ is invisible at the collider. However, the signature of dark sector particle $\psi^\pm$ can be looked at in the 
collider ~\cite{Bhattacharya:2015qpa,Bhattacharya:2017sml}. For example, $\psi^\pm$ can be pair produced through Drell-Yan processes. 
However, their decay can give interesting signatures. In particular, three body decay of $\psi^\pm$ can give interesting displaced 
vertex signatures. Here we assumed the mass of $X$-particles to be super heavy, namely $M_X\sim 10^{10}$ GeV. However, the mass scale 
of $X$-particles can be just above the EW scale if we assume resonant leptogenesis. In that case the decay of $X$-particles can give 
interesting signatures at collider. We will comeback to these issues in a future publication~\cite{future_draft}.

%%%%%%%%%%%%%%%%%%%%%%%%%%%%%%%%%%%%%%%%%%%%%%%%%%%%%%%%%%%%%%%%%%%%%%%%%%%%%

\end{document}